\documentclass[aps, prd, amsmath, floats, floatfix, twocolumn, nofootinbib, showpacs]{revtex4-1}
\usepackage{graphicx}
\usepackage{color}

\newcommand{\be}{\begin{eqnarray}}
\newcommand{\ee}{\end{eqnarray}}
\newcommand{\rar}{\rightarrow}

\begin{document}

\title{Can the supermassive objects at the centers of galaxies be traversable wormholes?\\ 
The first test of strong gravity for mm/sub-mm VLBI facilities}

\author{Cosimo Bambi}
\email{bambi@fudan.edu.cn}

\affiliation{Center for Field Theory and Particle Physics \& Department of Physics, Fudan University, 200433 Shanghai, China}

\date{\today}

\begin{abstract}
The near future mm/sub-mm VLBI experiments are ambitious projects aiming 
at imaging the ``shadow'' of the supermassive black hole candidate at the center 
of the Milky Way and of the ones in nearby galaxies. An accurate observation 
of the shape of the shadow can potentially test the nature of these objects and 
verify if they are Kerr black holes, as predicted by general relativity. However, 
previous work on the subject has shown that the shadows produced in other 
spacetimes are very similar to the one of the Kerr background, suggesting that 
tests of strong gravity are not really possible with these facilities in the near future. 
In this work, I instead point out that it will be relatively easy to distinguish black 
holes from wormholes, topologically non-trivial structures of the spacetime that 
might have been formed in the early Universe and might connect our Universe 
with other universes. 
\end{abstract}


\maketitle


It is thought that the center of every normal galaxy harbors a supermassive
black hole (BH) of $\sim 10^5 - 10^9$~$M_\odot$~\cite{kr}. Studies of the orbital
motion of gas and of individual stars point out the existence of dark concentrations
of mass too heavy, compact, and old to be clusters of non-luminous bodies,
as the cluster lifetime due to evaporation and physical collisions would be
shorter that the age of these systems~\cite{maoz}. The non-observation of 
thermal radiation emitted by the putative surface of these objects may be seen
as an indication for the absence of a normal surface and the presence of an
event horizon~\cite{eh1} (see however~\cite{eh2}). The widely accepted
interpretation is that all these supermassive objects are the Kerr BHs predicted
by general relativity, even if we do not yet have evidence that the geometry 
of the spacetime around them is really described by the Kerr solution~\cite{rev,pap}.

SgrA$^\star$ is a bright and very compact radio source at the center of the 
Milky Way. Its position coincides with the one of the supermassive BH
candidate of the Galaxy. At cm wavelengths, the structure of SgrA$^\star$ is
completely washed out by interstellar scattering. Observations of SgrA$^\star$
at mm wavelengths suggest that in the near future it will be possible to image 
the emission region around the supermassive BH candidate with very long
baseline interferometry (VLBI) techniques at sub-mm wavelengths~\cite{doe1}.
This can be possible for three favorable conditions. First, the interstellar
scattering reduces significantly at shorter wavelengths. Second, at sub-mm 
wavelengths, VLBI experiments can reach a resolution $\lambda/D \sim 
10$~$\mu$as, where $\lambda$ is the wavelength of the radiation and $D$ 
is the distance between different stations; such a resolution is of the order 
of the angular size of the gravitational radius of SgrA$^\star$. Third, the 
compact synchrotron emitting region of SgrA$^\star$ is expected to become 
optically thin at sub-mm wavelengths, thus allowing one to see the region around 
the supermassive object. Another good candidate for the observation of its
emission region is the supermassive object at the center of the galaxy M87~\cite{m87}.

General relativity makes clear predictions of the image of an optically thin emitting 
region around a BH. In particular, the image has a ``shadow'', a dark region over 
a brighter background~\cite{i-0,rohta}. While the intensity map of the image depends 
on the details of the accretion process and of the emission mechanisms, the 
boundary of the shadow is only determined by the metric of the spacetime, since it 
corresponds to the apparent image of the photon capture sphere as seen by a distant 
observer. The Event Horizon Telescope is an ambitious project with the main goal
to observe the shadow of SgrA$^\star$~\cite{doe2,eht}. The idea is to combine 
existing and planned mm/sub-mm facilities into a high resolution observatory.
Similar experiments include the Japanese VLBI Network (JVN)~\cite{jvn}, the 
Chinese VLBI Network (CVN)~\cite{cvn}, the Korean VLBI Network (KVN)~\cite{kvn}, 
the Chinese Space VLBI project, and the effort of joining these East Asia facilities 
together.

The possibility of testing the nature of supermassive BH candidates 
by observing the shape of their shadow has been already discussed in the literature. 
The idea was first explored in~\cite{i-1}, where the authors showed that the 
shadow of a Kerr BH and a of Kerr super-spinning object (a very compact object 
described by the Kerr solution with spin parameter $|a_*|>1$) are dramatically 
different, and that it would thus be relatively easy to distinguish the 
two cases with near future observations. 
However, super-spinning compact objects seem to be ruled out on general 
theoretical reasons, because they are impossible to create and, in any case, strongly
unstable~\cite{eb}. If we consider physically acceptable non-Kerr exotic 
compact objects or non-Kerr BHs in putative alternative theories of gravity, 
their shadow is not so different with respect to the one of a Kerr BH and only 
high precision observations can detect deviations from the Kerr predictions~\cite{i-2}. 
This is not a surprise. At first approximation, the boundary of the shadow is 
a circle, whose radius is determined by the distance and by the mass of 
the compact object. The first order correction comes from the spin of the 
compact object, i.e. the current-dipole term in a multipole moment expansion 
of the gravitational field. Deviations from the Kerr shadow correspond to 
second order deformations, as they are associated with higher order multipole 
moment terms. As the measurement of the spin parameter of SgrA$^\star$ 
and of the supermassive BH candidate at the center of M87 with the observation 
of their shadow is already out of reach for the Event Horizon Telescope and 
the East Asia projects, the possibility of testing the Kerr nature of these objects
seems to be even more unlikely.

The aim of this short paper is to show that this conclusion is not completely 
true. While the near future mm/sub-mm VLBI experiments will not be able 
to distinguish Kerr BHs from non-Kerr exotic compact objects or non-Kerr 
BHs, meaningful tests of the nature of SgrA$^\star$ are possible. In particular, 
it should be relatively easy to check if SgrA$^\star$ is actually a wormhole 
(WH)~\cite{wh}. WHs are topological structures of the 
spacetime connecting either two different regions of our Universe or two different
universes in Multiverse models. The non-observation of thermal radiation from
SgrA$^\star$ may be explained in the WH scenario, as WHs have no surface. WHs
at the center of galaxies may also bypass the puzzle of the existence of
BH candidates with $M \sim 10^9$~$M_\odot$ already at redshift $z \gtrsim 6$,
as they would be relics of the very early Universe. Studies to observationally 
distinguish BHs from WHs are already present in the literature~\cite{wh2,wh2b,wh3}
and they may represent the only way to test Multiverse models.

Before discussing specific cases, let us consider a generic static, spherically
symmetric, and asymptotically flat spacetime. Without loss of generality, its
line element can be written as
\be\label{eq-ds}
ds^2 = A(r) dt^2 - B(r) dr^2 - r^2 d\theta^2 - r^2 \sin^2\theta d\phi^2 \, ,
\ee
where $A(r)$ and $B(r)$ must reduce to 1 for $r \rar \infty$. Now we want to
determine the apparent size of the central object as seen by a distant observer;
that is, the photon impact parameter separating those photons falling onto the
object from the ones reaching a minimum distance and coming back to infinity.
Thanks to the spherical symmetry of the problem, we can restrict our discussion
to the equatorial plane $\theta = \pi/2$. As the metric coefficients in Eq.~(\ref{eq-ds})
do not depend on the $t$ and $\phi$ coordinates, there are two constants of motion,
the energy $E$ and the angular momentum $L$. From the Euler-Lagrangian
equations we find
\be
\dot{t} = \frac{E}{A(r)} \, , \quad \dot{\phi} = \frac{L}{r^2} \, .
\ee
$\dot{t}$ and $\dot{\phi}$ can then be plugged into $g_{\mu\nu} \dot{x}^\mu \dot{x}^\nu = 0$
to get
\be
\dot{r}^2 = \frac{1}{B(r)} \left[\frac{E^2}{A(r)} - \frac{L^2}{r^2}\right] \, .
\ee
If $\dot{r}^2$ vanishes before hitting the object, the photon reaches a turning point
at $\dot{r}^2 = 0$ and then comes back to infinity. In the opposite case, the
photon falls onto the object. The impact parameter is $b = L/E$ and the critical 
one separating captured and uncaptured photons is given by the solution of the 
system $\dot{r}^2 = 0$ and $\partial_r \dot{r}^2 = 0$, that is:
\be\label{eq-b}
&r^2 - b^2_{\rm crit} A(r) = 0 \, , & \nonumber\\
&2 b^2_{\rm crit} A^2(r) - r^3 A'(r) = 0 \, . &
\ee
$b_{\rm crit}$ corresponds to the value of the radius of the shadow as seen by a
distant observer. Let us note that $b_{\rm crit}$ depends only on $A(r)$, not on
$B(r)$.

In the Schwarzschild background, $A(r) = 1 - 2M/r$, and the solution of the 
system~(\ref{eq-b}) is
\be
b_{\rm Schwarzschild}/M = 3 \sqrt{3} \approx 5.196 \, .
\ee
Traversable WHs may have a quite different $A(r)$. There are many kinds of WHs
proposed in the literature, but a common WH geometry is the one with line
element (see e.g.~\cite{wh2b,wh3})
\be
ds^2 = e^{2\Phi(r)} dt^2 - \frac{dr^2}{1 - \Psi(r)} - r^2 d\theta^2 
- r^2 \sin^2\theta d\phi^2 \, ,
\ee
where $\Phi$ and $\Psi$ are, respectively, the redshift and the shape
function\footnote{The reader should note that the shape function is usually
defined in a different way in the literature of WHs.}. A common choice is
$\Phi = - r_0/r$, where $r_0$ is the WH throat radius and sets the scales of 
the system. $r_0$ is interpreted as the mass of the object in the Newtonian limit.
With this choice of the redshift factor, Eq.~(\ref{eq-b}) gives
\be
b_{\rm Wormhole}/r_0 = e \approx 2.718 \, ,
\ee
which is significantly smaller than the Schwarzschild prediction. If SgrA$^\star$
is a Schwarzschild BH, the diameter of the shadow should be 
\be
\theta_{\rm Schwarzschild} = \left(56 \pm 8\right) \; \mu{\rm as} \, ,
\ee
where the 14\% uncertainty comes from the uncertainty of the measurements
of its mass and its distance from us:
\be
&& M = \left( 4.31 \pm 0.38 \right) 10^6 \; M_\odot \;\; [23] \, , \nonumber\\
&& d  = 7.94 \pm 0.42 \; {\rm kpc} \;\; [24] \, .
\ee
For a WH, we have
\be
\theta_{\rm Wormhole} = \left(29 \pm 4\right) \; \mu{\rm as} \, .
\ee
The possibility that SgrA$^\star$ is an extremal Kerr BH (which is unlikely) 
slightly reduces this gap, as the expected angular size orthogonal to the direction 
of the spin would be (see e.g.~\cite{rohta})
\be
\theta_{\rm Kerr, \; a_*=1} = \left(48 \pm 7\right) \; \mu{\rm as} \, .
\ee
While the current uncertainty in the measurements of $M$ and $d$ are significant,
thus forbidding in general the possibility of measuring the spin parameter (assuming
the Kerr metric) or of distinguishing a Schwarzschild/Kerr BH from another object, in the 
case of WHs the difference in the expected apparent size is large enough to overcome 
these uncertainties. Spin parameter and deviations from Kerr can be 
measured/constrained without significant improvements of the measurements of
$M$ and $d$ only with high precision observations, as they can be inferred from the 
exact shape (rather than the size) of the shadow. This is out of reach for near future
experiments. Let us also note that the shadows of non-Kerr objects discussed so far in 
the literature differ from the Schwarzschild/Kerr one by a few percent~\cite{i-2}; that 
is, less than the uncertainty due to the measurements of $M$ and $d$.

\begin{figure*}
\begin{center}
\includegraphics[type=pdf,ext=.pdf,read=.pdf,width=5cm]{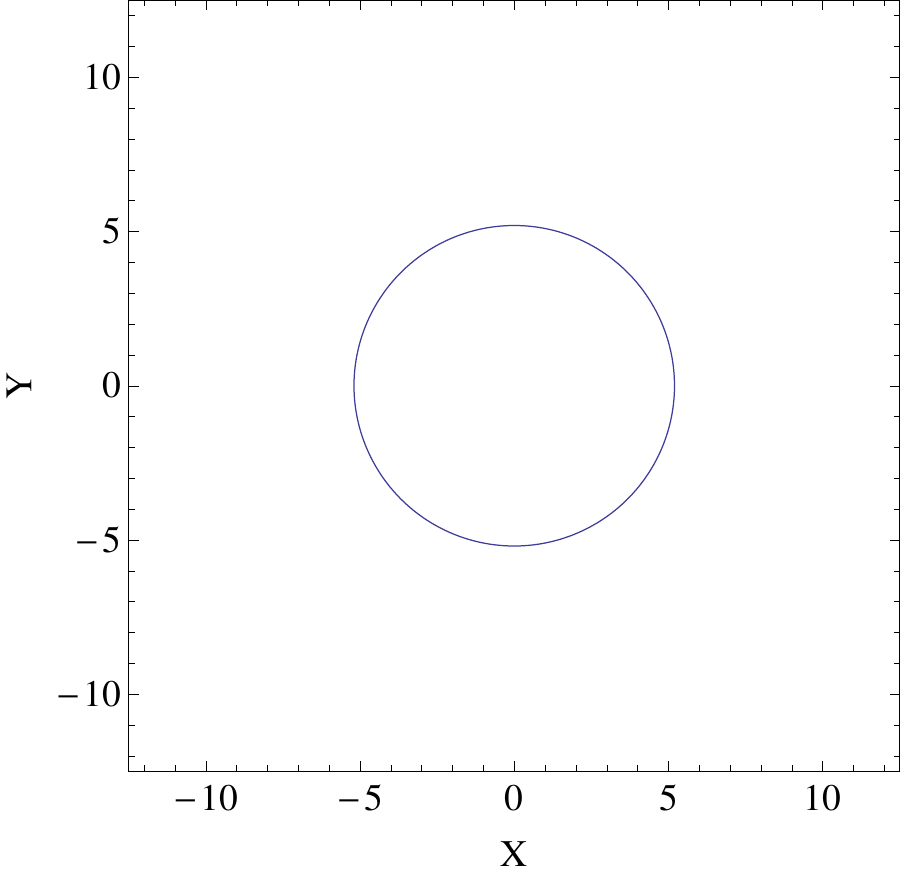} \hspace{0.5cm}
\includegraphics[type=pdf,ext=.pdf,read=.pdf,width=5cm]{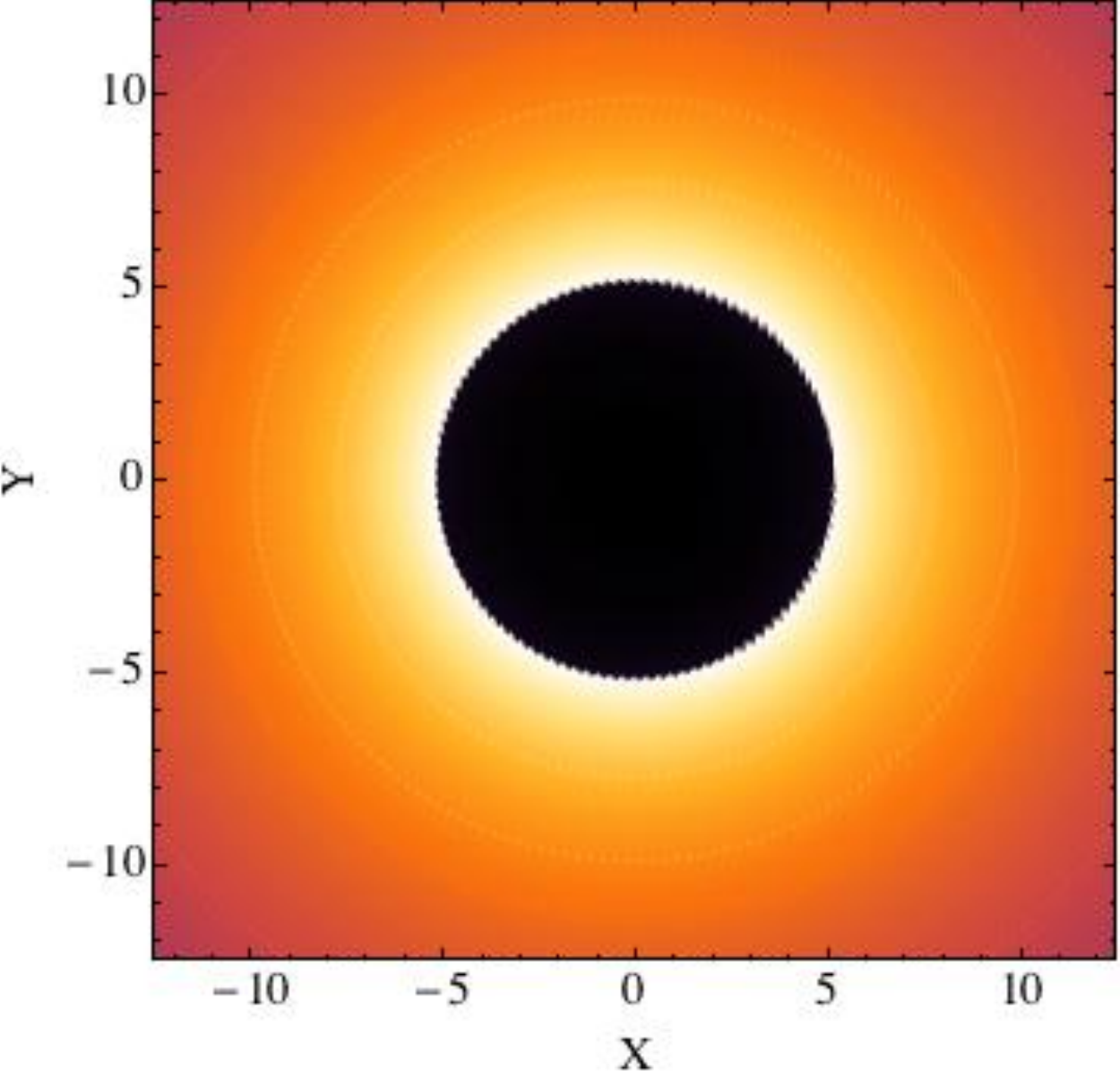} \hspace{0.5cm}
\includegraphics[type=pdf,ext=.pdf,read=.pdf,width=5cm]{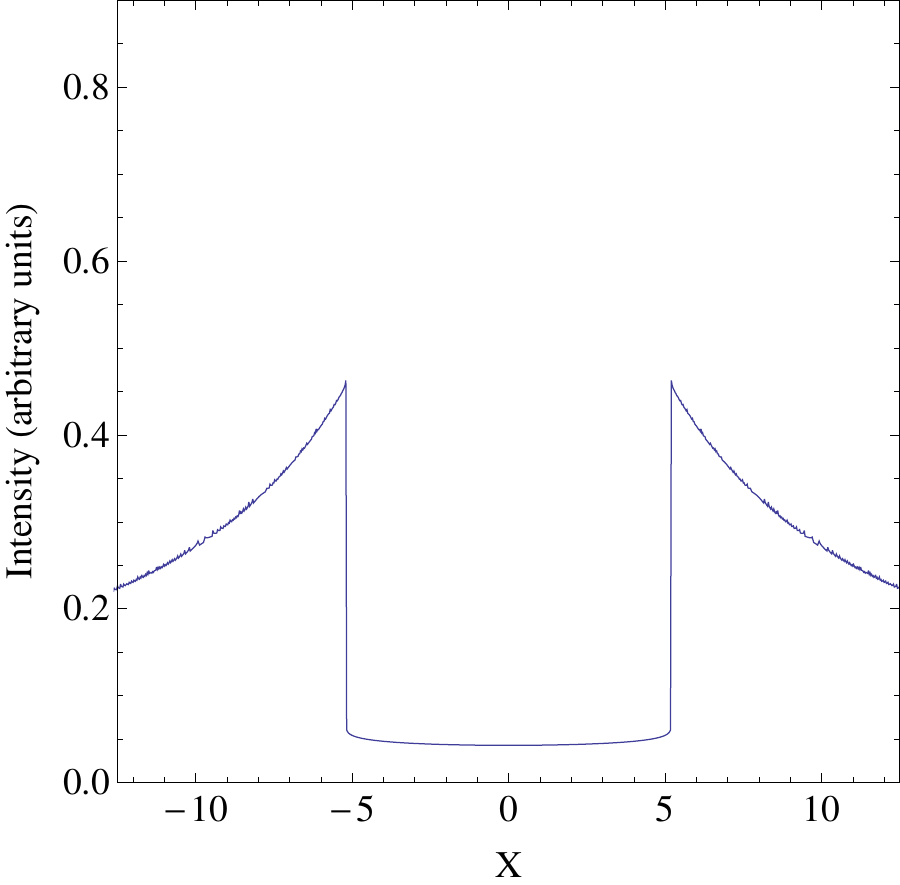}
 \end{center}
 \vspace{-0.5cm}
\caption{Schwarzschild black hole: contour of the black hole shadow (left panel),
image of an optically thin emission region surrounding the black hole (central 
panel), and its intensity variation along the $X$-axis (right panel). See the text for 
details.}
\label{f-bh}
\vspace{1cm}
\begin{center}
\includegraphics[type=pdf,ext=.pdf,read=.pdf,width=5cm]{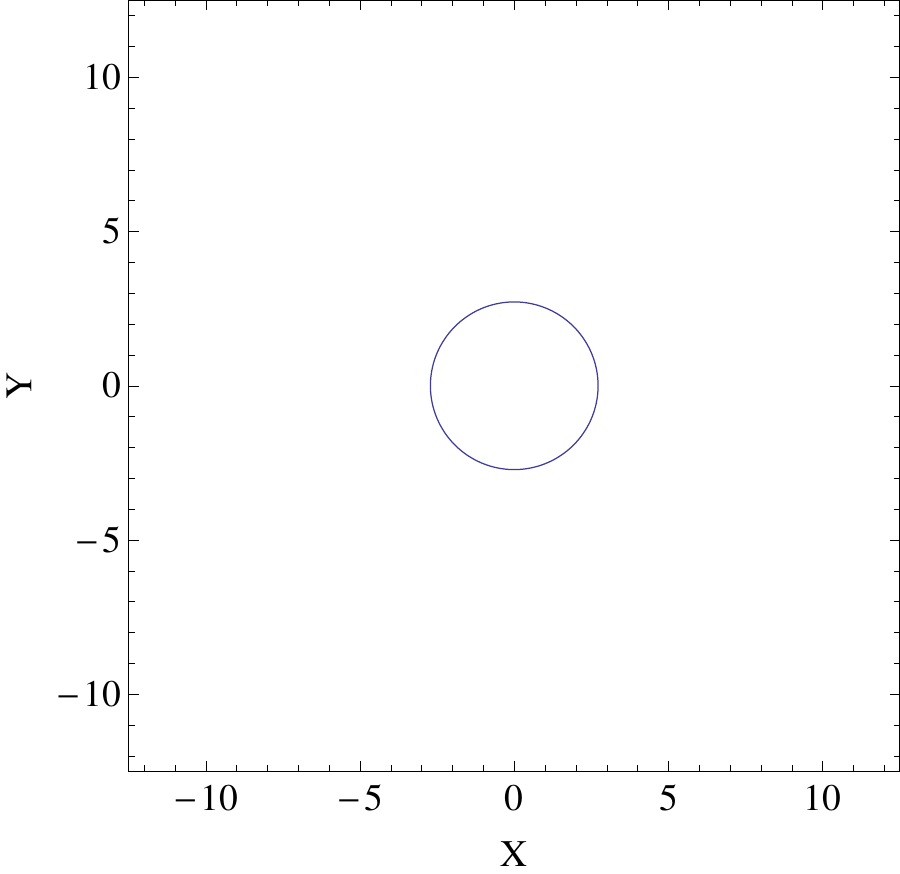} \hspace{0.5cm}
\includegraphics[type=pdf,ext=.pdf,read=.pdf,width=5cm]{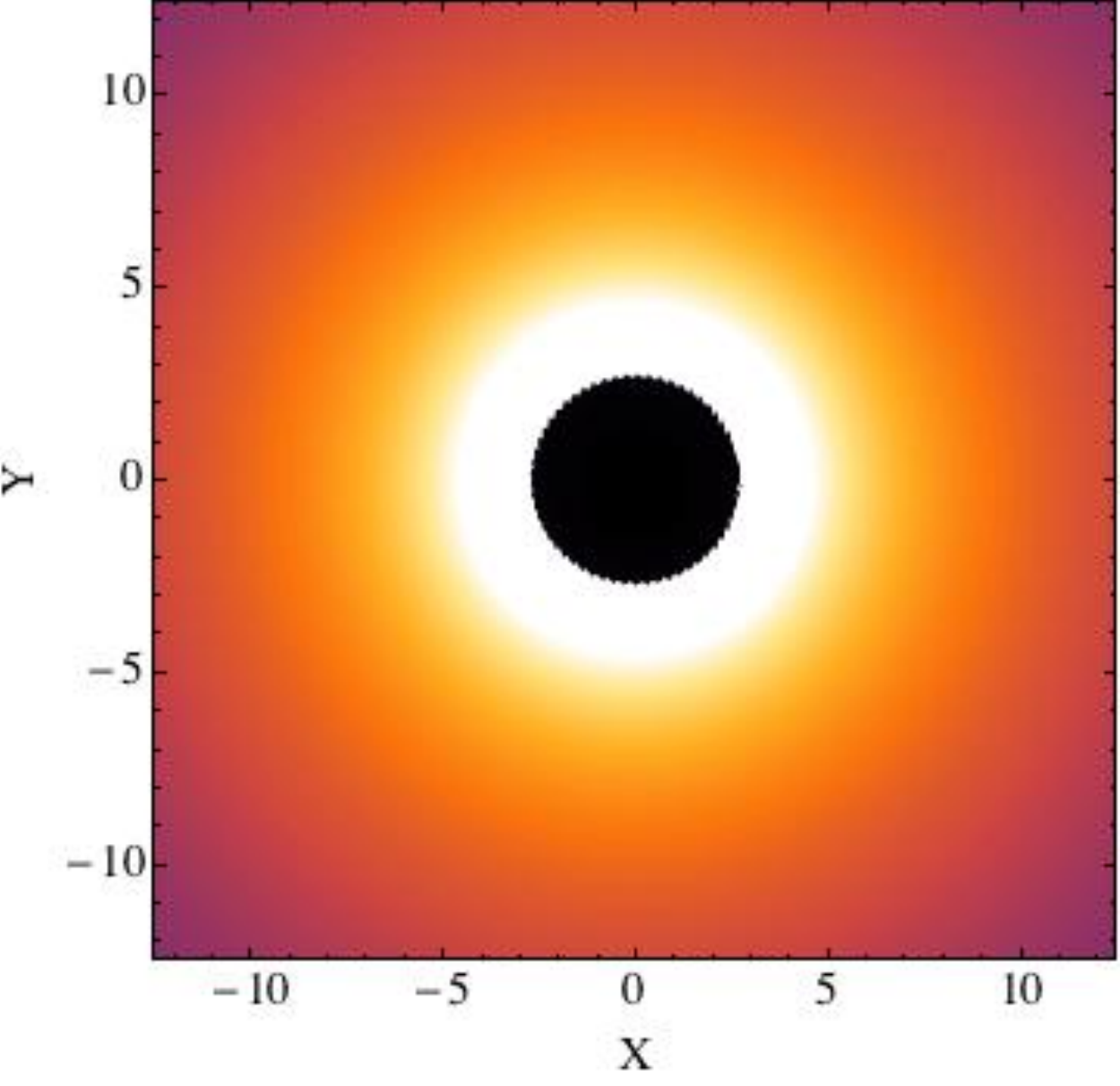} \hspace{0.5cm}
\includegraphics[type=pdf,ext=.pdf,read=.pdf,width=5cm]{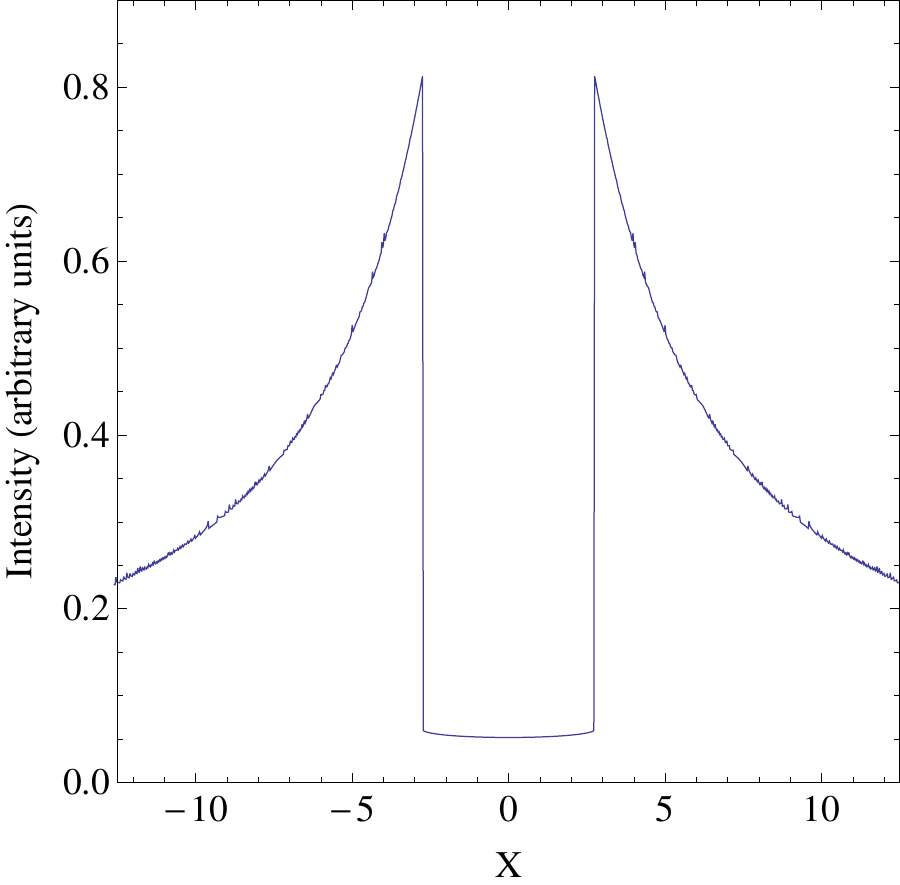}
 \end{center}
 \vspace{-0.5cm}
\caption{Traversable wormhole: contour of the wormhole shadow (left panel),
image of an optically thin emission region surrounding the wormhole (central 
panel), and its intensity variation along the $X$-axis (right panel). See the text for 
details.}
\label{f-wh}
\end{figure*}

The calculation of the intensity map of the image of the emitting region requires 
instead some assumptions about the accretion process and the emission mechanisms. 
The observed specific intensity at the observed photon frequency $\nu_{\rm obs}$ 
at the point $(X,Y)$ of the observer's image (usually measured in 
erg~s$^{-1}$~cm$^{-2}$~str$^{-1}$~Hz$^{-1}$) can be found integrating the 
specific emissivity along the photon path (see e.g.~\cite{ii})
\be\label{eq-I}
I_{\rm obs}(\nu_{\rm obs},X,Y) = \int_\gamma g^3 j (\nu_{\rm e}) dl_{\rm prop} \, ,
\ee
where $g = \nu_{\rm obs}/\nu_{\rm e}$ is the redshift factor, $\nu_{\rm e}$ is the
photon frequency as measured in the rest-frame of the emitter, $j (\nu_{\rm e})$ is the 
emissivity per unit volume in the rest-frame of the emitter, and $dl_{\rm prop}$ is the
infinitesimal proper length as measured in the rest-frame of the emitter. The
redshift factor can be evaluated from
\be
g = \frac{k_\alpha u^\alpha_{\rm obs}}{k_\beta u^\beta_{\rm e}} \, ,
\ee
where $k^\mu$ is the 4-momentum of the photon, $u^\mu_{\rm obs} = (1,0,0,0)$ is 
the 4-velocity of the distant observer, while $u^\mu_{\rm e}$ is the 4-velocity of 
the accreting gas emitting the radiation. Here, I consider the simple case of gas 
in free fall, which, in a static and spherically symmetric spacetime, reduces to
\be
u^t_{\rm e} = \frac{1}{A(r)} \, , \;\; 
u^r_{\rm e} = - \sqrt{\frac{1 - A(r)}{A(r) B(r)}} \, , \;\;
u^\theta_{\rm e} = u^\phi_{\rm e} = 0 \, .
\ee
For the photons, $k_t$ is a constant of motion and $k_r$ can be inferred
from $k_\alpha k^\alpha = 0$, that is:
\be
k_r = \pm k_t \sqrt{B(r) \left(\frac{1}{A(r)} - \frac{b^2}{r^2}\right)} \, ,
\ee
where the sign $+$($-$) is when the photon approaches (goes away from)
the massive object. $g$ is thus a function of $r$ and $b$. Unlike the calculation 
of $b_{\rm crit}$, which is determined only by $A(r)$, the intensity map
depends also on $B(r)$. Concerning the specific emissivity, we can
assume a very simple model in which the emission is monochromatic with 
rest-frame frequency $\nu_\star$ and a $1/r^2$ radial profile:
\be
j (\nu_{\rm e}) \propto \frac{\delta(\nu_{\rm e} - \nu_\star)}{r^2} \, ,
\ee
where $\delta$ is the delta function. Lastly, $dl_{\rm prop} = k_\alpha u^\alpha_{\rm e} 
d\lambda$ and, in our case, it reduces to
\be
dl_{\rm prop} = \frac{k_t}{g | k_r |} dr \, .
\ee
Integrating Eq.~(\ref{eq-I}) over all the observed frequencies, we get the 
observed photon flux
\be
F_{\rm obs} (X,Y) \propto \int_\gamma \frac{g^3 k_t dr}{r^2 | k_r |} \, .
\ee
The intensity map of the images for a Schwarzschild BH and a traversable
WH are shown in the central panels of, respectively, Figs.~\ref{f-bh} and \ref{f-wh}.
In the same figures, the right panels show the intensity along the X-axis.
In the case of the traversable WH, the calculations have used the shape 
function $\Psi = r_0/r$. It is also assumed that there is no or negligible emission 
of radiation coming from the other side of the WH.

{\it Conclusions ---}
The ``shadow'' of the supermassive BH candidate at the center of the Milky Way
and of the one in the galactic nucleus of M87 will be hopefully observed in the 
near future with mm/sub-mm VLBI facilities. As the shape of the shadow is only 
determined by the background metric around the compact object, these 
observations can potentially test the nature of these objects and verify if they are
Kerr BHs, as predicted by general relativity. However, the uncertainties in the 
measurements of their mass and distance is quite large. In the case of SgrA$^\star$, 
the expected size of the apparent image has an uncertainty of about 14\%,
while the deformations considered in the recent literature due to deviations from
the Kerr geometry are around a few percent. For the supermassive object at the 
center of M87, the predictions are even less clear. So, near future observations 
may not be able to distinguish Kerr BHs from other candidates. In this short work, 
I noted that the difference between the apparent image of Schwarzschild/Kerr 
BHs and WHs is instead quite significant, larger than the uncertainty coming 
from the measurements of $M$ and $d$, and thus relatively easy to test.

As a final remark, let us comment on possible wormholes with different 
geometries or the case of wormholes with non-vanishing spin. For a static and 
spherically symmetric spacetime, the size of the shadow is determined by 
$g_{tt}$, in the case of the wormholes discussed in this paper $g_{tt} = e^{-2r_0/r}$. 
While this form of $g_{tt}$ is a quite common choice in the literature, it is not 
unique. For instance, the one-way traversable wormholes discussed in~\cite{niko} 
have the exterior spacetime equivalent to the Schwarzschild/Kerr solution, and 
they are thus indistinguishable from the BHs of general relativity. The possibility 
of rotating wormholes should instead not change at all the conclusions of this 
work. The angular size would decrease for spinning solutions, so that the 
shadow of the Kerr family can always be distinguished from that of a spinning 
wormhole family. Moreover, rotating wormholes seem already to be to rule 
out by current X-ray data~\cite{wh3}.


\begin{acknowledgments}
I thank Rohta Takahashi for useful discussions and suggestions.
This work was supported by the Thousand Young Talents 
Program and Fudan University.
\end{acknowledgments}


\end{document}